\documentclass[pra,aps,amsmath,amssymb,lengthcheck,superscriptaddress,floatfix]{revtex4}
\usepackage{graphicx}% Include figure files
\usepackage{dcolumn}% Align table columns on decimal point
\usepackage{bm}% bold math
\usepackage[usenames,dvipsnames]{color}
\usepackage{psfrag}
 
\newcommand{\comm}[1]{}

\begin{document}
%\pacs{\tt Not for circulation}

%\title{Every Thing You Always Wanted to Know About the Bose-Hubbard  Dimer \\ But were afraid to ask }
\title{Quantum Criticality in a Bosonic Josephson Junction}
%Quantum counterpart of a dynamical bifurcation in the BH model}

\author{P. Buonsante}
\affiliation{Dipartimento di Fisica, Universit\`a degli Studi di Parma, V.le G.P. Usberti n.7/A, 43100 Parma, Italy}
\affiliation{Istituto Nazionale di Ottica-CNR (INO-CNR) and European Laboratory for Non-Linear Spectroscopy (LENS) Via N. Carrara 1, I-50019 Sesto Fiorentino, Italy}
\author{R. Burioni}
\affiliation{Dipartimento di Fisica, Universit\`a degli Studi di Parma, V.le G.P. Usberti n.7/A, 43100 Parma, Italy}
\author{E. Vescovi}
\affiliation{Dipartimento di Fisica, Universit\`a degli Studi di Parma, V.le G.P. Usberti n.7/A, 43100 Parma, Italy}
\author{A. Vezzani}
 \affiliation{Centro S3, CNR Istituto di Nanoscienze	, via Campi 213/a, 41100 Modena, Italy}
\affiliation{Dipartimento di Fisica, Universit\`a degli Studi di Parma, V.le G.P. Usberti n.7/A, 43100 Parma, Italy}
\date{\today}

\begin{abstract}
In this paper we consider a bosonic Josephson junction described by a two-mode Bose-Hubbard model, and we thoroughly analyze a quantum phase transition occurring in the system in the limit of infinite bosonic population. We discuss the relation between this quantum phase transition and the dynamical bifurcation occurring in the spectrum of the Discrete Self Trapping equations describing the system at the semiclassical level.
In particular, we identify five regimes depending on the strength of the effective interaction among bosons, and study the finite-size effects arising from the finiteness of the bosonic population.
We devote a special attention to the critical regime which reduces to the dynamical bifurcation point in the thermodynamic limit of infinite bosonic population. Specifically, we highlight an anomalous scaling in the population imbalance between the two wells of the trapping potential, as well as in two quantities borrowed from Quantum Information Theory, i.e. the entropy of entanglement and the ground-state fidelity.
Our analysis is not limited to the zero temperature case, but considers thermal effects as well.
\end{abstract}
\maketitle
\section{Introduction}
Several years ago it was suggested that two noninteracting condensates trapped in a double well trap at zero temperature would give rise to the bosonic analog of a Josephson junction \cite{Javanainen_PRA_57_3164}. Further studies of such a bosonic Josephson junction followed, where the effect of boson interaction was taken into account in a semiclassical framework based on the so-called discretized Gross-Pitaevskii equations \cite{Smerzi_PRL_79_4950,Raghavan_PRA_60_R1787,Smerzi_PRA_61_063601,Franzosi_IJMPB_9_943}. Different interesting regimes and behaviors were predicted  for the system depending on the value of the effective interaction among the bosons in each of the two wells of the trapping potential. These include the so-called macroscopic self-trapping phenomenon and a dynamical bifurcation corresponding to a localization transition in the ground-state of the system  \cite{attr_rep_map}.
Just a few years after the above mentioned theoretical work, brilliant experiments were carried out with ultra-cold atomic gases where the predicted Josephson oscillations, macroscopic nonlinear self-trapping and dynamical bifurcation were observed \cite{Albiez_PRL_95_010402,Gati_JPB_40_R61,Zibold_PRL_105_204101}.

The quantum version of the discrete Gross-Pitaevskii equations --- known as discrete self-trapping (DST) equations  \cite{Eilbeck_PhysicaD_16_318} among the nonlinear community --- was also widely investigated \cite{Scott_PLA_119_60,Bernstein_Nonlinearity_3_293,Milburn_PRA_55_4318}. In particular, a significant amount of attention  \cite{Cirac_PRA_57_1208,Spekkens_PRA_59_3868,Ho_JLTP_135_257,Jack_PRA_71_023610,Buonsante_PRA_72_043620,Shchesnovich_PRA_78_023611,Zin_EPL_83_64007,Lee_PRL_102_070401,Buonsante_PRA_82_043615,Mazzarella_PRA_83_053607} was directed to the quantum counterpart of the abovementioned dynamical transition from a localized to a delocalized regime, which, in this paper will be clearly recognized, as a quantum phase transition. 

In particular we carry out a detailed analysis of the ground state of the system which, at the quantum level, is described by a two-mode Bose-Hubbard model. We identify five regimes depending on the effective bosonic interaction. The central, critical regime corresponds to the dynamical bifurcation occurring in the solutions of the time-independent DST equations at the semiclassical level. Actually, this regime shrinks with increasing bosonic population, and reduces to the semiclassical critical point in the limit of infinite population which, in the present setting, has the role of a thermodynamic limit \cite{Buonsante_PRA_84_061601}. Focusing on the  ``standard'' order parameter, i.e. the population imbalance of the two wells, as well as on  quantities borrowed from quantum information theory, such as the entropy of entanglement and the ground-state fidelity \comm{citare}, we calculate the critical exponents characterizing the quantum phase transition in the thermodynamic limit. Moreover, we study the finite-size scaling effects, evidencing, within the critical regime, the anomalous scaling of the considered quantities as a function of the total boson population.
Our analysis is not restricted to the ground-state of the system, i.e. to the zero-temperature case. We also consider finite temperatures, and evidence  a different scaling of the quantum and thermal fluctuations of the  population imbalance in the critical regime.

The plan of the paper is as follows. In Section \ref{model} we describe the quantum Hamiltonian employed in the  description of the two-mode system, and discuss its relation with similar models, as well as with the semiclassical DST equations; In Section \ref{Sle} we illustrate how the low-lying spectrum of the quantum Hamiltonian can be equivalently investigated by means of a Schr\"odinger-like equation for a fictitious particle on a one-dimensional domain, and observe that the semiclassical DST equations naturally emerge from a small-oscillation approach around the minima of the potential energy.
Section \ref{HarmAppr} discusses the so-called Gaussian regimes, where the  Schr\"odinger-like equation can be safely approximated by a  Schr\"odinger equation proper, featuring a harmonic potential part. The limits of validity of such an approximation are analyzed in Section \ref{lov}, where further non-Gaussian regimes are discussed. Some more technical issues concerning the strong-coupling regimes are deferred to Appendix \ref{Apr}.
The critical regime is detailedly analyzed in Sec. \ref{crS}, whereas Sec. \ref{QITi} is devoted to a discussion of the behavior of two interesting indicators borrowed from Quantum Information Theory, the entropy of entanglement and the ground-state fidelity. The thermal fluctuations of the order parameter are investigated in Sec. \ref{thermS}, while Sec \ref{concl} contains our conclusions and perspectives. 

\section{The Model}
\label{model}
The Hamiltonian of the system reads
\begin{equation}
\label{H}
H = \frac{U}{2} H_{\rm int}- \frac{\Omega}{2} H_{\rm kin}-\frac{v}{2} H_{\rm bias}
\end{equation}
where
\begin{align}
H_{\rm int} &= \left(a_1^\dag a_1^\dag  a_1 a_1+a_2^\dag a_2^\dag  a_2 a_2\right) \nonumber \\
\label{HU}
& = \left[ n_1(n_1-1)+n_2(n_2-1) \right],\\
\label{HO}
H_{\rm kin} & = \left(a_1^\dag a_2 + a_2^\dag a_1\right)\\
\label{HB}
H_{\rm bias} & = \left(n_1- n_2\right)
\end{align}
and the lattice operators $a_j^\dag$, $a_j$ and $n_j= a_j^\dag a_j$ respectively create, destroy and count bosons at lattice site $j$. The operators in Eqs.~\eqref{HU} and \eqref{HO} account for interactions among bosons at the same site and for their kinetic energy, respectively. The relative importance of these terms is controlled by the interaction strength $U$ and the hopping amplitude $\Omega>0$. The operator $H_{\rm bias}$ allows for a local energy offset between the two sites, breaking the mirror symmetry of the system.
 
Notice that Hamiltonian \eqref{H} commutes with the total number operator, $N=n_1+n_2$, which therefore can be considered as a constant for all practical purposes. Owing to this fact, the inclusion of a further term of the form $\Gamma H_{\rm dip} =\Gamma n_1\,n_2$, accounting for an interaction between bosons at different lattice produces a Hamiltonian of the same class as $H$ \cite{Javanainen_PRA_60_2351}. Note indeed that $ H_{\rm dip} = \frac{1}{2}(N^2-N-H_{\rm int})$. 

Also notice that a simple mapping exists between the spectral  features of Hamiltonian \eqref{H} for attractive $U<0$ and repulsive $U>0$ interactions. This can be appreciated by observing that $D \left[\frac{U}{2} H_{\rm int}+\frac{v}{2} H_{\rm bias} - \frac{\Omega}{2} H_{\rm kin} \right] D^{-1} = -\left[-\frac{U}{2} H_{\rm int} -\frac{v}{2} H_{\rm bias} - \frac{\Omega}{2} H_{\rm kin} \right]  $, where $D=e^{i\, \pi (n_1-n_2)}$ is a unitary transform, \cite{Franzosi_PRA_63_043609}. 

Hamiltonian \eqref{H} can be studied by assuming that the state of the system is well approximated by a trial wavefunction of the form
\begin{equation}
\label{su2}
|\Phi\rangle\!= \frac{1}{\sqrt{N!}} \left(\sqrt{\frac{1+z}{2}} e^{i \frac{\varphi}{2}} a_1^\dag + \sqrt{\frac{1-z}{2}} e^{-i \frac{\varphi}{2}} a_2^\dag\right)^N\! |0\rangle
\end{equation}
where $-1\leq z \leq 1 $ and $\varphi$ are two real dynamical variables describing the population imbalance and the macroscopic phase difference between the two wells \cite{Wright_PhysD_69_18,Raghavan_PRA_60_R1787}. Eq. \eqref{su2} is a su$(2)$ coherent state, as discussed in Ref.~\cite{Buonsante_PRA_72_043620}.
The dynamics of the above variables is determined by the expectation value of the trial state on Eq. \eqref{H}, which
plays the role of a semiclassical Hamiltonian,
\begin{equation}
\label{mfH}
{\cal H} = \langle \Phi |H|\Phi\rangle = \frac{\Omega N}{2} \left(\frac{\gamma}{2} z^2 -\varepsilon \,z - \sqrt{1-z^2} \cos\varphi\right)
\end{equation}
where $\gamma = \frac{UN}{\Omega}$ and $\varepsilon = \frac{v}{\Omega}$ are the effective parameters of the model \cite{Smerzi_PRL_79_4950}.
Specifically, the dynamics is dictated by the so-called  {\it Discrete Self-Trapping} equations \cite{Eilbeck_PhysicaD_16_318}  ensuing from Eq.~\eqref{mfH}, and the corresponding fixed-point equations are known to exhibit  bifurcations at $|\gamma|^{2/3}-1=|\varepsilon|^{2/3}$ \cite{Buonsante_JPB_37_S229,Shchesnovich_PRA_78_023611}.
The manifestation of the semiclassical bifurcations in the features of the quantum system has been repeatedly reported in the literature.  For instance the emergence of structures in the spectrum of the quantum problem \eqref{H} at the same energies as the semiclassical fixed-point solutions has been investigated in Refs.~\cite{Aubry_PRL_76_1607,Cirac_PRA_57_1208,Franzosi_PRA_63_043609,Karkuszewski_EPJD_21_251,Buonsante_JPB_37_S229}

In the following we will be mainly concerned with the 
mirror-symmetric case, $\varepsilon=0$, for which the
bifurcation corresponds to a dynamical transition at $\gamma=-1$ 
in the ground-state
of Hamiltonian \eqref{mfH}. For $\gamma \geq -1$ this is symmetric,
$z=0$, while for $\gamma<-1$, due to the nonlinear nature of the
dynamical equations,  the mirror symmetry of the Hamiltonian is
spontaneously broken, $z=\pm \sqrt{1-\gamma^{-2}}$. 

Although the ground-state of the quantum Hamiltonian \eqref{H}  is always 
symmetric, $\langle n_1\rangle=\langle n_2\rangle$, a crossover in its 
internal structure can be recognized in the vicinity of the
mean-field critical value. Several indicators have been employed to
highlight this crossover: energy gaps \cite{Cirac_PRA_57_1208,Pan_PLA_339_403,Oelkers_PRB_75_115119},
fluctuations in the population imbalance
\cite{Spekkens_PRA_59_3868,Ho_JLTP_135_257,Shchesnovich_PRA_78_023611,Zin_EPL_83_64007,Mazzarella_PRA_83_053607},
structure of the occupation number distribution \cite{Spekkens_PRA_59_3868,Ho_JLTP_135_257,Javanainen_PRL_101_170405,Lee_PRL_102_070401,Mazzarella_PRA_83_053607}
fidelity \cite{Oelkers_PRB_75_115119}, entanglement entropy
\cite{Hines_PRA_71_042303,Pan_PLA_339_403,Fu_PRA_74_063614,JuliaDiaz_PRA_81_023615,Mazzarella_PRA_83_053607}, coherence visibility
\cite{Ho_JLTP_135_257,Mazzarella_PRA_83_053607}, generalized purity
\cite{Viscondi_PRA_80_013610}.
Refs. \cite{Jack_PRA_71_023610,Buonsante_PRA_72_043620,Oelkers_PRB_75_115119,Buonsante_PRA_82_043615,Buonsante_PRA_84_061601}
consider similar issues on systems comprising more than two sites. 

We mention that Hamiltonian \eqref{H} can be mapped onto a Lipkin-Meshkov-Glick (LMG) model with vanishing anisotropy parameter. Some of the results we illustrate in the following have been obtained  in that context making use of numerical or analitical techniques different from ours\cite{Botet_PRL_49_478,Dusuel_PRL_93_237204,Dusuel_PRB_71_224420,Ribeiro_PRE_78_021106,Kwok_PRE_78_032103}.

\section{The Schr\"odinger-like equation for the low-lying
  eigenstates}
\label{Sle}
Under suitable conditions, the eigenvalue equation for Hamiltonian \eqref{H}, $H |\Psi\rangle = E |\Psi\rangle$ is very satisfactorily approximated by a Schr\"odinger-like equation for a fictitious particle on a one-dimensional domain. Several slightly different versions of this approach have been employed the literature \cite{Javanainen_PRA_60_2351,Javanainen_PRA_60_4902,Franzosi_IJMPB_9_943,Franzosi_PRA_63_043609,Jaaskelainen_PRA_71_043603,Shchesnovich_PRA_78_023611,Zin_PRA_78_023620,Zin_EPL_83_64007,JuliaDiaz_PRA_81_023615,JuliaDiaz_PRA_81_063625}. All of them are based on the expansion of the eigenstate on the Fock basis of the site occupation numbers
\begin{equation}
\label{eigval}
|\Psi\rangle = \sum_{\nu=0}^N c_\nu |N-\nu\rangle_1 \otimes |\nu\rangle_2,\quad |\nu\rangle_j = \frac{\left(a_j^\dag\right)^{\nu}}{\sqrt{\nu!}} |0\rangle,
\end{equation}
where $|0\rangle$ is the vacuum state, $a_j |0\rangle=0$. 
Introducing the normalized population imbalance of the Fock state relevant to expansion coefficient $c_\nu$
\begin{equation}
\label{zFock}
z_\nu = \frac{(N-\nu)-\nu}{N} = 1- \frac{2\nu}{N}
\end{equation}
the eigenvalue equation reads
\begin{align}
&\frac{U N^2}{4} z_\nu^2\, c_\nu -\frac{\Omega N}{2} \left[\sqrt{\frac{1+z_\nu}{2}\left(\frac{1-z_\nu}{2}-\frac{1}{N}\right)} \,c_{\nu-1}\right. \nonumber\\
\label{eeFs}
&
\left. \sqrt{\frac{1-z_\nu}{2}\left(\frac{1+z_\nu}{2}-\frac{1}{N}\right)}\,
  c_{\nu+1}\right]+\frac{v N}{2} z_\nu\, c_\nu = \bar E \,c_\nu
\end{align}
where the "rescaled energy'' $\bar E = E+U\left(2 N-N^2\right)/4-\Omega/2=E+\frac{\Omega}{2}[\frac{\gamma}{2}(N-2)-1]$
differs from the original eigenvalue by a unimportant population-dependent constant.
Now, for large boson populations $N$, both the square roots appearing in in Eq.~\eqref{eeFs} and the coefficients $c_{\nu\pm 1}$ can be expanded in powers of $N^{-1}$. Note indeed that the latter can be seen
as the values that a continuous function $\psi$ takes on at the points of the  grid mesh in the interval $[-1, 1]$ defined by $z_\nu$ with $\nu=0,1,2,\cdots,N$. Specifically, if we set $c_{\nu \pm 1} =\sqrt{\frac{2}{N}}\, \psi(z_{\nu\pm 1})= \sqrt{\frac{2}{N}}\, \psi(z_\nu \mp 2N^{-1})$  \cite{entdc} and retain  only the lowest order terms in the expansion of the interaction and kinetic term in the eigenvalue equation, we obtain the following Schr\"odinger-like equation  \cite{Zin_EPL_83_64007}:
\begin{equation}
\label{ceq1}
\frac{\Omega N}{2} \left[-\frac{2}{N^2} \frac{d}{d z} \sqrt{1-z^2}
  \frac{d}{d z} + V_\varepsilon(z)\right] \,\psi(z) = \bar E\,\psi(z)
\end{equation}
where $z\in [-1,\, 1]$ is the continuum limit of the mesh grid and
 the effective potential is \cite{Shchesnovich_PRA_78_023611}
\begin{equation}
\label{pot1}
V_\varepsilon(z) = \frac{\gamma}{2} z^2-\sqrt{1-z^2}+\varepsilon\, z
\end{equation}

Owing to the $N^{-2}$ coefficient in front of the derivative part in
Eq.~\eqref{ceq1}, for large populations the low-lying solutions of
such equation will be strongly localized in the vicinity of the minima
of the potential energy, Eq.~\eqref{pot1}. This suggest that the
Schr\"odinger-like equation \eqref{ceq1} can be further simplified by
introducing the (small) deviation from the minima of Eq. \eqref{pot1}.
Note that the minimization of Eq.~\eqref{pot1} is equivalent to the
determination of the lowest-energy stationary solution of the Discrete
Self-Trapping equations arising from the mean-field Hamiltonian
\eqref{mfH} \cite{Buonsante_PRA_84_061601}. This is easily verified in the mirror symmetric case,
$\varepsilon=0$, where, up to the fourth order in the small
displacement from the minimum point(s), $u=z-z_\gamma$,  equation \eqref{ceq1} is equivalent to
\begin{equation}
\label{ceq2}
\left[ -\frac{d^2}{d u^2}+ A\,u^2+B\,u^3+C\,u^4\right]
\varphi(u) = {\cal E}\varphi(u)
\end{equation}
with $\psi(z) = \varphi(z-z_\gamma)$ and
\begin{widetext}
% \begin{equation}
% \begin{array}{|c|c|c|c|c|c|}
% \hline
% & z_\gamma & A & B & C & {\cal E} \\
% \hline
% \scriptstyle{\gamma\geq -1} & 0 & \frac{N^2 (\gamma+1)}{4} & 0 & \frac{N^2}{16} &
% \frac{\bar E N}{\Omega}+\frac{N^2}{2} \\
% \hline
% \scriptstyle{\gamma< -1} & {\pm\sqrt{\frac{\gamma^2-1}{\gamma^2}}} & \frac{N^2 \gamma(1-\gamma^2)}{4} & \frac{N^2 \gamma^4(1-\gamma^2)}{4} & \frac{N^2 \gamma^5(1-5\gamma^2)}{16} &
% \frac{E N}{\Omega}+\frac{N^2}{2} \\
% \hline
% \end{array}
% \end{equation}
\begin{equation}
\label{coeff}
\begin{array}[b]{|c|c|c|c|c|c|}
\hline
& z_\gamma & A & B & C & {\cal E} \\[.5em]
\hline
\gamma\geq -1 & 0 &\displaystyle{ \frac{N^2 (\gamma+1)}{4}} & 0 & \displaystyle{\frac{N^2}{16}} &
\displaystyle{\frac{\bar E N}{\Omega}+\frac{N^2}{2}} \\[1.3em]
\hline
\gamma< -1 & {\pm\sqrt{\displaystyle{1-\frac{1}{\gamma^2}}}} &
\displaystyle{\frac{N^2 \gamma^2 (\gamma^2-1)}{4}} &
\mp \displaystyle{\frac{N^2 \gamma^5\sqrt{\gamma^2-1}}{4}} & \displaystyle{\frac{N^2 \gamma^6(5\gamma^2-4)}{16}} &
\displaystyle{\frac{N^2(1+\gamma^2)}{4}-\gamma\frac{\bar E N}{\Omega}} \\[1.2em]
\hline
\end{array}
\end{equation}
\end{widetext}

\section{Harmonic regime}
\label{HarmAppr}

We note that the leading order in Eq. \eqref{ceq2} is almost everywhere
the second, except at the mean-field critical point $\gamma=-1$ where
the first non-vanishing term is the fourth. Therefore, for the lowest 
energy levels, away from criticality and for sufficiently large populations, we can let
$B=C=0$ and Eq. \eqref{ceq2} reduces to
the Schr\"odinger equation for a harmonic oscillator, whose discrete
spectrum is 
\begin{equation}
\label{hoE}
{\cal E}_n = \sqrt{4 A} \left(n+\frac{1}{2}\right) = \frac{1}{\sigma_\gamma^2} \left(n+\frac{1}{2}\right) 
\end{equation}
where
\begin{equation}
\label{sigma}
\sigma_\gamma^2 = \frac{1}{2 \sqrt{A}}= \left\{
\begin{array}{ll}
\displaystyle \frac{1}{N\sqrt{\gamma+1}} & \gamma>-1 \\[1em]
\displaystyle \frac{1}{N |\gamma|\sqrt{\gamma^2-1}} & \gamma<-1 
\end{array}
\right.
\end{equation}

Actually, for $\gamma < -1 $  Eq. \eqref{ceq2} describes the small
oscillations about either of the two equivalent minima of the
effective potential, $z_\gamma =\pm
\sqrt{1+\gamma^-2}$. Therefore a generic solution of Eq. \eqref{ceq1}
is well approximated by a superposition of one harmonic solution for
each well. The symmetry of the problem dictates that these
superpositions have a definite parity,
\begin{equation}
\label{supG}
\psi_n^\pm(z)=\frac{\varphi_n(z-|z_\gamma|)\pm
  \varphi_n(z+|z_\gamma|)}{\sqrt{2}}
\end{equation} 
Also, since in this approximation the two localized functions  $\varphi_n(z\pm
z_\gamma)$ have a vanishing overlap, both $\psi_n^\pm(z)$ have an
energy very close to that in Eq. \eqref{hoE}.
In particular, the ground-state of the system $\psi_0(z)$ is well approximated
by a symmetric superposition $\psi_0^+(z)$ of two Gaussian functions of the form
\begin{equation}
\label{Gauss}
\varphi_0(u) = \frac{1}{(\sigma_\gamma \sqrt{2  \pi})^{1/2}} e^{-\frac{u^2}{4\sigma_\gamma^2}} 
\end{equation}
with $\sigma_\gamma^2$ given in (the lower of) Eq. \eqref{sigma}.

For $\gamma>-1$ the effective potential has a single minimum at
$z_\gamma=0$, and the problem can be mapped exactly onto a quantum
harmonic oscillator, as far as the width of the eigenfunctions of the
latter problem do not exceed the interval in which the harmonic term
dominates over higher-order terms. The energy spectrum is again given by
Eq. \eqref{hoE} and, in particular, the ground-state is $\psi_0(z) =
\varphi_0(z)$, Eq. \eqref{Gauss}, with
$\sigma_\gamma^2$ given  in (the upper of)
Eq. \eqref{sigma}. 

The most natural physical quantity which is in
principle measurable, is given by the population imbalance
corresponding to the  operator \cite{Zin_EPL_83_64007} %\comm{altre ref??}
in the original two-site problem:
\begin{equation}
\label{pio}
\hat z = \frac{a_1^\dag a_1-a_2^\dag a_2}{N},
\end{equation}
Its value and fluctuations are easily evaluated on the ground state of the system
\begin{align}
\label{ez}
\langle \hat z \rangle &= \int ds\, z |\psi_0(z)|^2 = 0
\end{align}
and \cite{intN}
\begin{align}
\label{ez2}
\langle \hat z^2 \rangle &= \int ds\, z^2 |\psi_0(z)|^2 \approx
\sigma_\gamma^2+z_\gamma^2 
\end{align}
where $\langle \cdot \rangle$ denotes the expectation value on the ground state of the system. Moreover one can also consider the expectation value $\langle \cdot \rangle_n$ on the $n$th excited state obtaining:
\begin{align}
\label{ez2n}
\langle \hat z^2 \rangle_n &= \int ds\, z^2 |\psi_n(z)|^2 \approx
\sigma_\gamma^2(2n+1)+z_\gamma^2 
\end{align}

\begin{figure}
\includegraphics[width= \columnwidth]{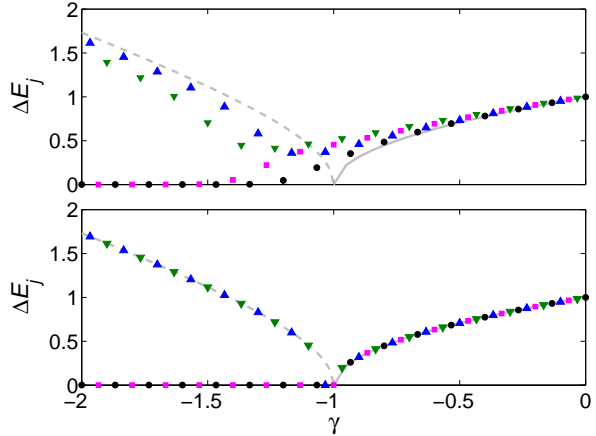}
\caption{\label{en_gaps} (color online) Low lying energy gaps for $N=50$ (upper
  panel) and $N=5000$ (lower panel). Gray thick lines represent the
  analytic results in Eqs. \eqref{DeltaO} and
\eqref{DeltaE}. Specifically, the dashed line is
  $\Omega \sqrt{\gamma^2-1}$ and the  solid line is
  $\Omega \sqrt{\gamma+1}$. Symbols represent numerical data. Specifically,
circles, upward triangles, squares and downward triangles are  $\Delta
E_n$ with $n=1,2,3,4$, respectively.}
\end{figure}

The fluctuations of the population imbalance have been investigated by
several authors,
\cite{Spekkens_PRA_59_3868,Ho_JLTP_135_257,Shchesnovich_PRA_78_023611,Mazzarella_PRA_83_053607},
and analytical expressions for this quantity have been reported in
Refs. \cite{Spekkens_PRA_59_3868,Shchesnovich_PRA_78_023611,Mazzarella_PRA_83_053607}. 
In both Gaussian regimes the (low lying) energy spectrum of
Eq. \eqref{ceq2} is well approximated by Eq. \eqref{hoE}. Since $E_n =
\bar E_n -\frac{\Omega}{2}[\frac{\gamma}{2}(N^2-2N)-1]$, and $\bar E_n$ depends on ${\cal
  E}_n$ as described in table \eqref{coeff}, it is easy to
calculate the lowest energy gaps in the spectrum of $H$, $\Delta E_n= E_n-E_{n-1}$. 
As we mentioned earlier, for $\gamma_1\ll \gamma \ll \gamma_2$ each energy level is
(almost) twofold degenerate, so that we have
\begin{equation}
\label{DeltaO}
\Delta E_{2 n+1} \approx \left\{
\begin{array}{ll}
0, & \gamma_1\ll \gamma \ll \gamma_2 \\
\Omega \sqrt{\gamma+1},& \gamma_3\ll \gamma \ll \gamma_4
\end{array}
\right.
\end{equation}
and
\begin{equation}
\label{DeltaE}
\Delta E_{2 n} \approx \left\{
\begin{array}{ll}
\Omega \sqrt{\gamma^2-1}, & \gamma_1\ll \gamma \ll \gamma_2 \\
\Omega \sqrt{\gamma+1},& \gamma_3\ll \gamma \ll \gamma_4
\end{array}
\right.
\end{equation}
In Fig. \ref{en_gaps} we compare Eqs. \eqref{DeltaO} and
\eqref{DeltaE} with numerical data for $\Delta E_n$. The plots clearly
show that the analytic approximations become more accurate with
with increasing boson population.
  
\section{Limits of validity and regimes}
\label{lov}
In the present section we describe the limits of validity of the
harmonic solutions described above and the different regimes
taking over at different values of the effective parameter $\gamma$.
A crucial assumption in passing from Eq.~\eqref{eeFs} to
Eq.~\eqref{ceq1} is that the coefficients $c_\nu$ can be regarded as
continuous functions of the population imbalance of the Fock state
defined in Eq.~\eqref{zFock}. This may be problematic for eigenstates
in the mid-spectrum of  Hamiltonian~\eqref{H}, but is very reasonable for
states close to the extrema of the same spectrum.
However, the continuous approach can prove ill posed even for the ground
state of the system. 
It is clear that, for the  continuous limit of
Eq.~\eqref{ceq1} to be well defined, the Gaussin width 
$\sigma_\gamma$ must be much larger than
distance between two neighboring points in the discrete
Eq.~\eqref{eeFs}, namely $2\, N^{-1}$. This means  $\gamma \gg
\gamma_1 \sim - \sqrt{N} $ for $\gamma<-1$ and $|\gamma| \ll
\gamma_4 \sim N^2$ for $\gamma>-1$ \cite{Javanainen_PRA_60_2351}.

We observe that the same estimates can be obtained by 
imposing that the expansion parameter in the strong coupling regimes
be actually small, as discussed in some detail in Sec. \ref{Apr}.
This ensures that for $\gamma \gg \gamma_1$ and $\gamma \ll \gamma_4$
the first-order perturbative description of the ground state applies.

The parameter range $\gamma_1 \ll \gamma \ll \gamma_4$ where the continuous
approach applies can be divided into three regimes.
For $\gamma_1 \ll \gamma \ll \gamma_2 $ the ground state of the system is
well approximated by a symmetric superposition of two Gaussians.
 On the other hand, for $\gamma_3 \ll \gamma \ll \gamma_4$ the ground state is well approximated by a single Gaussian.

These two regimes are separated by an intermediate {\it critical}
regime $\gamma_2 < \gamma  < \gamma_3 $, surrounding the mean-field critical point $\gamma_{\rm c}=-1$, where the higher order terms in Eq.~\eqref{ceq2} cannot be
neglected and the problem is not Gaussian any more. The
boundaries of this regime can be obtained from a comparison between the leading and
next-to-leading term in the perturbative expansion of the effective
potential.  A reasonable estimate for the lower bound $\gamma_2<\gamma_{\rm c}$
can be obtained either by imposing $A\,\sigma_\gamma^2 \gg |B|
\sigma_\gamma^3 $ or by requiring that the distance
between the two equivalent harmonic minima, $2\sqrt{1-\gamma^{-2}}$, is
much larger than the width of the relevant Gaussian ground states
$\sigma_\gamma$. In both cases one finds $\gamma_{\rm c}-\gamma_2\sim N^{-2/3}$.
Similarly, the
upper bound $\gamma_3>\gamma_{\rm c}$ is obtained by requiring $A\,\sigma_\gamma^2 \gg C
\,\sigma_\gamma^4$, which results once again into
$ \gamma_3 - \gamma_{\rm c} \sim  N^{-2/3}$ \cite{Shchesnovich_PRA_78_023611,Zin_PRA_78_023620}.
%An alternate route to the same estimate of the boundaries of the
%critical region, based on the properties of the energy spectrum of Hamiltonian \eqref{H}, is discussed in Appendix \ref{crA}.
As discussed in Sec. \ref{crS},  the energy gaps in the critical regime depend on the system population, $\Delta E \sim N^{-1/3}$, at variance with what happens in
the Gaussian regimes. This provides a $N-$dependent cutoff to the
vanishing of $\Delta E$ at $\gamma=\gamma_{\rm c}$ predicted  in Eqs. \eqref{DeltaO} and \eqref{DeltaE}.

In summary, one can recognize five different regimes depending on the
value of the effective parameter $\gamma$. 
 For $\gamma \ll \gamma_1$ and $\gamma\gg \gamma_4$ the continuous approximation of Section
\ref{Sle} does not apply, but the system can be analyzed by resorting
to first-order perturbative theory, as discussed in Appendix
\ref{Apr}. For $\gamma_1 \ll \gamma \ll \gamma_2$ the continuous
approximation applies, and the ground-state of the system is a
symmetric superposition of two Gaussians centered at the equivalent harmonic
minima of the effective potential. For $\gamma_2 \ll \gamma \ll
\gamma_3$ the system is critical, in a sense that will be clarified
shortly. Note that in the limit of large populations the critical
regime reduces to the mean-field critical point. For $\gamma_3 \ll \gamma \ll \gamma_4$ the ground-state is a
Gaussian centered at the unique minimum of the effective potential,
$z_\gamma=0$.
Note that in estimating the location of the boundaries $\gamma_{1-4}$
we only considered the dependence on the (large) bosonic population,
and neglected other numerical factors.

\section{Critical regime}
\label{crS}
According to our previous analysis, as the effective parameter
$\gamma$ approaches the mean-field critical value $\gamma_{\rm c}=-1$,
the system exhibits the hallmarks of a phase transition, where the 
minimum $z_\gamma$ of the effective potential, table \eqref{coeff}, 
%\begin{equation}
%\label{zg}
%z_\gamma = \left\{
%\begin{array}{ll}
%\pm \sqrt{1-\gamma^{-2}} & \gamma < \gamma_{\rm c}\\
%0 & \gamma>\gamma_{\rm c}
%\end{array}
%\right.
%\end{equation} 
plays
the role of the order parameter. Note that this quantity is in
principle measurable as the absolute value of the population imbalance
operator  $z_\gamma = \langle|\hat z|\rangle$. 
The fluctuations of this order parameter
\begin{equation}
\label{zf}
\langle |\hat z|^2\rangle-\langle |\hat z|\rangle^2 = \sigma_\gamma^2 
\end{equation}
coincide with the Gaussian
width of Eq. \eqref{sigma}.

The behavior for $z_\gamma$ and $\sigma_\gamma^2$ described in table \eqref{coeff} and Eq. \eqref{zf}, respectively, holds true in the mirror-symmetric case, $v=\varepsilon=0$.
It proves useful to study behaviour of the order parameter in
the presence of a vanishing energy offset between the sites of the
dimer, $|v| \ll1$. According to our discussions about the Gaussian
regimes, the value of the order parameter will coincide with the
absolute minimum of the effective potential which, owing to the
presence of a non-vanishing $\varepsilon$, is now non degenerate for
any value of $\gamma$. The minimization of Eq. \eqref{pot1} then gives
\begin{equation}
\label{ze}
\langle |\hat z|\rangle_\varepsilon = \left\{
\begin{array}{ll}
\langle |\hat z|\rangle_{\varepsilon=0} -\frac{\varepsilon}{\gamma(\gamma^2-1)} & \gamma<-1 \\
\langle |\hat z|\rangle_{\varepsilon=0}+\frac{\varepsilon}{\gamma+1} & \gamma>-1
\end{array}
\right.
\end{equation}
where only linear terms in $\varepsilon$ have been taken into account.
Eq. \eqref{ze} allows us to introduce a sort of susceptibility 
\begin{equation}
\label{chiz}
\chi_z = \frac{d}{d \epsilon} \langle |\hat z|\rangle_\varepsilon = \left\{
\begin{array}{ll}
\frac{1}{|\gamma|(\gamma^2-1)} & \gamma<-1 \\[0.5em]
\frac{1}{\gamma+1} & \gamma>-1
\end{array}
\right.
\end{equation}
Equations \eqref{coeff}, \eqref{sigma}, \eqref{zf}, \eqref{chiz} provide
some critical exponents for the system. As $\gamma$ approaches its
critical value $\gamma_{\rm c}=-1$, the order parameter vanishes (from
below) as $\langle|\hat z|\rangle \sim (\gamma_{\rm
  c}-\gamma)^{\alpha_z}$, its fluctuations diverge as
$\langle |\hat z|^2\rangle-\langle |\hat z|\rangle^2 \sim
|\gamma-\gamma_{\rm c}|^{-\alpha_\sigma}$
and its susceptibility diverges as $\chi_z \sim |\gamma-\gamma_{\rm
  c}|^{-\alpha_\chi}$, where
\begin{equation}
\alpha_z = \alpha_\sigma = \frac{1}{2}, \qquad \alpha_\chi = 1 
\end{equation}

The above discussion rigorously applies in the limit of infinite
population, which plays the role of an effective thermodynamic limit.
Indeed, as we observe in Section \ref{lov}, only in this limit does
the critical regime shrink to a single point corresponding to the
 critical value.
For any finite populations there exists a finite region surrounding
the critical point where the Gaussian results do not apply. 
In this region the quartic term dominates and Eq. \eqref{ceq1} can be approximated as
\begin{equation}
\label{ceq3}
  \left[-\frac{d^2}{d z^2} + C z^4\right] \psi_n(z) = {\cal E}_n \psi_n(z)
\end{equation}
where $C$ is listed in the upper row of table \eqref{coeff}.
It is easy to prove that in this critical regime the  eigenfunctions of Eq. \eqref{ceq3} have form 
\begin{equation}
\label{cr_psi}
\psi_n(z) = N^{1/6} \phi_n(z N^{1/3}), 
\end{equation}
where $\phi_n(\zeta)$ is the
solution of the population-independent problem
\begin{equation}
\label{ceq4}
  \left[-\frac{d^2}{d \zeta^2} +\frac{\zeta^4}{16}\right]
  \phi_n(\zeta) = \bar {\cal E}_n \phi_n(\zeta)
\end{equation}
and the eigenvalue corresponding to $\psi_n(z)$ is ${\cal E}_n
=N^{2/3} \bar {\cal E}_n$. 
This result allows us to estimate the behaviour of the (low-lying)
energy gaps in the spectrum of Hamiltonian \eqref{H} in the critical
regime. We get $\Delta E_n \sim N^{-1/3} $, where we
used the definition of ${\cal E}$ in table \eqref{coeff} and the fact that the 
spectrum of problem \eqref{ceq4} does not
depend on $N$. 
\begin{figure}
\includegraphics[width= \columnwidth]{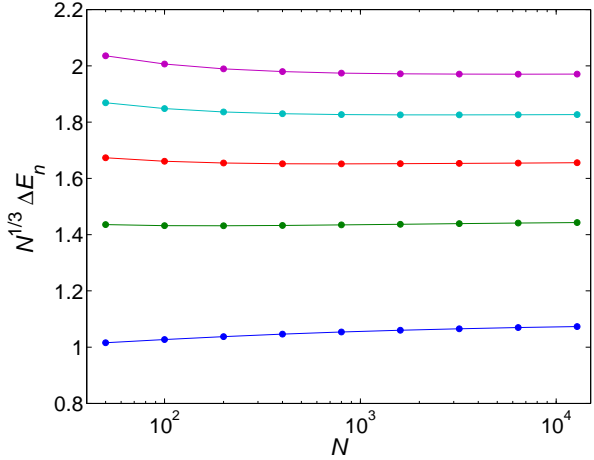}
\caption{\label{en_gaps_cr} (color online) Dependence on the population of the first
  few energy gaps $\Delta E_n$ at $\gamma =-1$ (from bottom to top,
  $n=1,2\cdots,5$). The data points are numerically determined values,
  the solid lines are just guides to the eye.}
\end{figure}
This estimate, whose correctness demonstrated in
Fig. \ref{en_gaps_cr}, provides an alternate route for the determination of
the boundaries of the critical region. Indeed, these can be assessed
by requiring that the prediction in the Gaussian regimes,
Eqs. \eqref{DeltaO} and \eqref{DeltaE}, are of the
same order as that in the critical regime, i.e. $N^{1/3}$. This
gives once again $\gamma_{\rm c}-\gamma_2\sim N^{-2/3}$ and
$\gamma_3-\gamma_{\rm c} \sim N^{-2/3}$.

As to the fluctuations of the order parameter, particularizing Eq. \eqref{cr_psi} to the ground-state of the system, $n=0$, we get
\begin{equation}
\label{zfc}
\langle |\hat z|^2\rangle-\langle |\hat z|\rangle^2 = N^{-2/3}\frac{\int d\zeta \,\zeta^2\,|\phi_0(\zeta)|^2}{\int d\zeta \,|\phi_0(\zeta)|^2}
\end{equation}
Within a standard scaling approach it is reasonable to assume that there 
exists a correlation length, dictating the effective size of the system,
which diverges at the critical point
\begin{equation}
\label{sca2}
{\cal N}_\gamma \sim |\gamma-\gamma_{\rm c}|^{-\xi}.
\end{equation}
Any physical quantity should depend on the boson population $N$ only through
the ratio $N/{\cal N}_\gamma$, so that it should be
\begin{equation}
\label{sca1}
\langle |\hat z|^2\rangle-\langle |\hat z|\rangle^2 = N^\tau
f\left(\frac{N}{{\cal N}_\gamma}\right)
\end{equation}
The comparison with Eqs. \eqref{sigma}, \eqref{zf} and \eqref{zfc} then allows us to
conclude that 
\begin{equation}
\label{sca3}
\tau = -\frac{2}{3},\qquad \xi = \frac{3}{2} 
\end{equation}
\begin{figure}
\includegraphics[width=8cm]{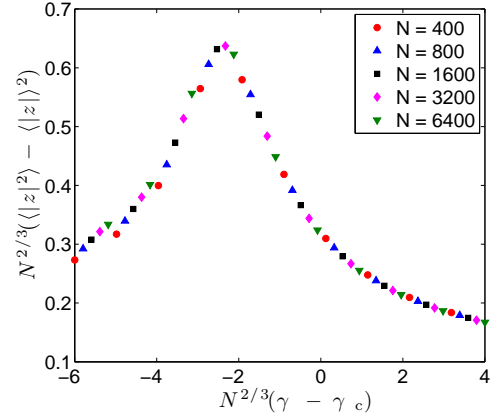}
\caption{\label{scaling_fl} (color online) Scaling of the fluctuations of the  (absolute value of) the population imbalance operator. }
\end{figure}
The correctness of the scaling assumption in
Eqs. \eqref{sca1}-\eqref{sca3} is demonstrated in
Fig. \ref{scaling_fl}, where a nice data collapse is observed for a
wide range of boson populations.
We mention that -- in the context of the LMG model -- the critical exponents in \eqref{sca3} were determined numerically in Ref.~\cite{Botet_PRL_49_478} and making use of an analytic approach different from ours in Refs.~\cite{Dusuel_PRL_93_237204,Dusuel_PRB_71_224420,Ribeiro_PRE_78_021106}.

We conclude by remarking that the rescaled energies $ \bar {\cal E}_n $ and the expectation values for the rescaled problem \eqref{ceq4} can be estimated by 
applying the Bohr-Sommerfeld quantization rule. This gives
\begin{equation}
\label{ceq5}
\int_{-2{\cal E}_n^{1/4}}^{2{\cal E}_n^{1/4}} \sqrt{\bar {\cal E}_n -\frac{\zeta^4}{16}}d\zeta=\pi(n+\frac 1 2), \qquad n\in \mathbb{N},
\end{equation}
which provides quite satisfactory  estimates of the energy levels of the system,
\begin{equation}
\label{ceq6}
E_n= \Omega C_1 N^{-1/3} (n+\frac 1 2)^{2/3}
\end{equation}
where $C_1$ is a suitable constant. Within the same approximation the mean square displacement of the eigenfunction $\phi_n(\zeta)$ can be evaluated as well. In particular, for the simclassical orbit of energy $\bar {\cal E}_n$, one gets
\begin{align}
\int \phi_n(\zeta)^2 \zeta^2\, d\zeta &\approx \int \zeta^2 \delta(p^2 + \frac{\zeta^4}{16} -\bar {\cal E}_n) \,dp \, d\zeta \nonumber\\
\label{ceq7}
&= C_2 (n+\frac 1 2)^{2/3}
\end{align}
where $C_2$ is a suitable constant.

\section{QIT Indicators}
\label{QITi}

As we mention above, quantities borrowed from quantum information
theory have been employed to investigate the the incipient quantum
phase transition between the localized and delocalized ground-state of
the system. Owing to the particularly simple structure of the system, 
some of these quantities turn out to be entirely equivalent to usual
indicators. This is the case of the generalized purity of
Ref.~\cite{Viscondi_PRA_80_013610}, which basically coincides with the
coherence visibility considered in
Refs.~\cite{Ho_JLTP_135_257,JuliaDiaz_PRA_81_023615,Mazzarella_PRA_83_053607},
i.e. the off-diagonal term of the one-body density matrix.
Similarly, the Fisher information of
Ref.~\cite{Mazzarella_PRA_83_053607} is equivalent to the population
imbalance fluctuations considered in several earlier works
\cite{Spekkens_PRA_59_3868,Ho_JLTP_135_257,Shchesnovich_PRA_78_023611,Zin_EPL_83_64007}
as well as in the present paper.

The fidelity and entanglement entropy considered in
Refs.~\cite{Oelkers_PRB_75_115119} and
\cite{Mazzarella_PRA_83_053607}, respectively, appear to be less
trivial quantities \cite{fident}. The former is  nothing but the overlap
between two ground-states of the system corresponding to slightly
different values of the parameter driving the transition
$\langle\Psi_\gamma|\Psi_{\gamma+\kappa}\rangle$ \cite{Zanardi_PRE_74_031123}.
Since this quantity is almost everywhere extremely close to unity, a
more convenient indicator is provided by the so-called {\it fidelity
  susceptibility} \cite{Cozzini_PRB_76_104420}
\begin{equation}
\label{fidS1}
\chi_{\rm f} = \lim_{\kappa \to 0}
\frac{1-\langle\Psi_\gamma|\Psi_{\gamma+\kappa}\rangle}{\kappa^2}
= \frac{1}{2} \frac{d^2}{d \kappa^2} \langle\Psi_\gamma|\Psi_{\gamma+\kappa}\rangle\Big|_{\kappa=0}
\end{equation} 

In a bipartite system, the entropy of entanglement of a pure state $|\Psi\rangle$ is
given by $-{\rm Tr} (\rho_j \log \rho_j)$, where $\rho_j$ is the
reduced density matrix of subsystem $j$, obtained from the full density
matrix $|\Psi\rangle\langle\Psi|$ after tracing over the states of the
other subsystem \cite{Bennet_PRA_53_2046}. In the case under investigation the two
subsystem are of course the two sites composing the Bose-Hubbard
dimer and, owing to number conservation, the entropy of entanglement
of an eigenstate of Hamiltonian \eqref{H} is
\begin{equation}
\label{entent}
S_{\rm e} = -\frac{1}{\log_2 (N+1)} \sum_{\nu=0}^N |c_\nu|^2 \log_2 |c_\nu|^2 
\end{equation} 
where the $c_\nu$'s are the expansion coefficients in
Eq.~\eqref{eigval} and we introduced a normalization factor. This
quantity has been considered in a few earlier works on the
Bose-Hubbard dimer. In the presence of repulsive interactions it was used to investigate the precursor
of  the Mott insulator--superfluid quantum phase transition occurring on
infinite lattices \cite{Hines_PRL_67_013609}. In attractive systems it
was employed for the characterization of {\it
  Schr\"odinger-cat}-like states in the presence of a small energy
offset $v$ between the two sites of the system \cite{JuliaDiaz_PRA_81_023615},  and in the study of
the quantum counterpart of the mean-field transition for small finite boson populations \cite{Mazzarella_PRA_83_053607}.

These two indicators have been considered for the LMG model \cite{Kwok_PRE_78_032103,Vidal_JSM_P01015} which, as we mention, can be mapped onto Hamiltonian \eqref{H}. However, the natural decomposition of the system necessary for  the calculation of the entanglement entropy is quite different in the spin and boson context, and the results are not readily comparable. As to the ground-state fidelity, a comparison will be made with the results in Ref.~\cite{Kwok_PRE_78_032103}.

Based on the results sketched in the previous sections and in the appendices, we are able to
give an analytic description of the behaviour of the fidelity
susceptibility  and entropy of entanglement, Eqs. \eqref{fidS1} and
\eqref{entent}.

\begin{figure}
\includegraphics[width= \columnwidth]{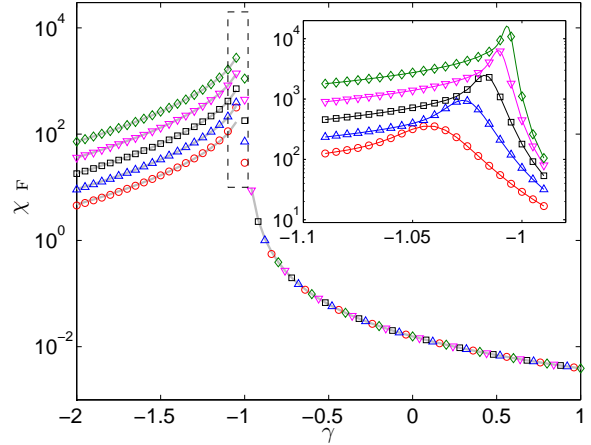}
\caption{\label{fidF1} (color online) The analytic expression obtained for the fidelity susceptibility
  in the Gaussian regimes is compared with the numerical results
  for the same quantity. The circles, upward triangles, squares,
  downward triangles and diamonds are obtained from exact
  diagonalization of Eq. \eqref{H} for 
systems containing $500$, $1000$, $2000$, $4000$ and $8000$ bosons,
respectively. In all cases we used Eq. \eqref{fidS1} with $\kappa=10^{-4}$.
The thick gray lines represent Eq. \eqref{fidS2} for the same boson
populations. Notice how $\chi_{\rm F}(\gamma)$ depends on the boson
population only for $\gamma<\gamma_{\rm c}$. The inset contains a
magnification of the region enclosed by the dashed black frame, where
Eq. \eqref{fidS2} does not apply. The thin solid lines are guides to
the eye.   }
\end{figure}

Let us start with the former quantity. As we discuss in Section
\ref{Sle}, in the Gaussian regimes
the ground-state of the system is strictly related to the ground state
of a harmonic oscillator. Specifically, for
$\gamma_1\ll\gamma\ll\gamma_2$ the ground state is 
 well approximated  by a symmetric superposition , $\psi_0^+(z)$, of
two Gaussian functions of square width $\sigma_\gamma^2 = \left(N
  |\gamma|\sqrt{\gamma^2-1}\right)^{-1}$ centered at $z_\gamma = \pm
\sqrt{1-\gamma^{-2}}$, as described by Eqs. \eqref{supG} and \eqref{Gauss}.
Likewise, for $\gamma_3ll\gamma\ll\gamma_4$ the ground state is well
described by a single
Gaussian $\varphi_0(z)$ of square width  $\sigma_\gamma^2 = \left(N
  \sqrt{1+\gamma}\right)^{-1}$ centered at $z_\gamma=0$, as described
by Eq. \eqref{Gauss}. Straightforward calculations result in
\begin{equation}
\label{fidS2}
\chi_{\rm F}(\gamma) \!=\! \left\{
\begin{array}{ll}
\frac{N}{8|\gamma|^3\sqrt{\gamma^2-1}}+ \frac{(2\gamma-1)^2}{16 \gamma^2(\gamma^2-1)^2}& \gamma_1 \!\ll\! \gamma \!\ll\! \gamma_2 \\
\frac{1}{64 \left(\gamma+1\right)^2}& \gamma_3 \!\ll\! \gamma \!\ll\! \gamma_4
\end{array}
\right.
\end{equation}
whose correctness is demonstrated in Fig. \ref{fidF1}.
It is worth noticing that Eq. \eqref{fidS2} predicts a dependence on
the boson population only in the lower Gaussian regime $\gamma_1\ll
\gamma \ll \gamma_2$, whereas for $\gamma_3\ll
\gamma \ll \gamma_4$ the fidelity susceptibility turns out to be $N-$independent.
 This can be explained by the fact that in the latter regime
the small variation $\kappa$ in the effective parameter affects only the
width of the single Gaussian representing the ground-state of the
system. Conversely, for $\gamma_1\ll \gamma \ll \gamma_2$, the change
affects both the
centers and the widths of the two Gaussians comprising the ground state
of the system.
Also notice that Eq. \eqref{fidS2} predicts a divergence in the
fidelity susceptibility at the mean-field critical point,
$\gamma=\gamma_{\rm c}$. However, for any finite $N$,
there is a small region surrounding such point where the Gaussian
approximations do not apply. This is demonstrated in the inset of
Fig. \ref{fidF1}, which clearly shows that for finite populatons $\chi_{\rm F} (\gamma)$
features a finite peak in the vicinity of $\gamma_{\rm c}$. The
larger the population, the sharper and the closer to the critical
point is the peak. 

In order to study the behavior of $\chi_{\rm F}(\gamma)$ in the
critical region it proves useful to exploit Eq. \eqref{ceq3}. 
As it is discussed in Ref. \cite{You_PRE_76_022101}, the limit in Eq.~\eqref{fidS1} can be carried out exactly through a perturbative expansion involving the entire spectrum of the problem at a given value of the driving parameter $\gamma$. Considering the harmonic term in Eq. \eqref{ceq3} as a perturbation over the quartic term, we get
\begin{equation}
\label{fidS3}
\chi_F(\gamma_{\rm c}) = \frac{N^{4/3}}{16}\sum_{n\neq 0} \left[\frac{\int d\zeta
    \,\zeta^2\, \phi_0(\zeta) \phi_n(\zeta) }{\bar {\cal E}_0-\bar
  {\cal E}_n }\right]^2 
\end{equation}
where the eigenfunctions $\phi_n(\zeta)$ and eigenvalues $\bar {\cal E}_n$ 
are dfined by \eqref{cr_psi} and \eqref{ceq4}.
Rigorously speaking the $\phi_n(\zeta)$ provides a good
approximation for an eigenstate of the original problem,
Eq. \eqref{H}, only for sufficiently small values of the quantum
number $n$. However, only the first few terms in the sum appearing in
Eq. \eqref{fidS3} are expected to provide a significant contribution,
owing to the localization of the ground state $\phi_0(\zeta)$ and
the energy gap appearing in the denominator.

\begin{figure}
\includegraphics[width= \columnwidth]{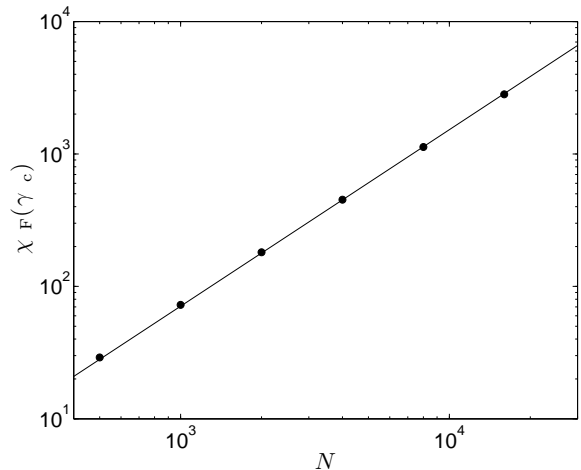}
\caption{\label{fid_cr} Scaling of the fidelity susceptibility at the
  mean-field critical value. The data points are obtained from exact
  diagonalization of system containing $N=500 \cdot 2^n$ bosons, with
  $n=0,\cdots,5$. The thin solid line corresponds to $7.1\cdot 10^{-3}\,N^{4/3}$.
}
\end{figure}

The $N^{4/3}$ behavior predicted in Eq. \eqref{fidS3} and confirmed by
the numerical results in Fig. \ref{fid_cr} can be
recognized as the {\it superextensive} divergence of the fidelity
susceptibility which, according to Ref. \cite{Cozzini_PRB_76_104420}, is the hallmark of
a quantum phase transition. The same exponent was estimated numerically in Ref.~\cite{Kwok_PRE_78_032103}. Also, our Eq.~\eqref{fidS2} agrees with a similar result in the same paper, although we obtain a different behavior of the sub-leading terms.

Let us now turn to the entropy of entanglement. Recalling that the continuous version of Eq.~\eqref{entent} is
\begin{equation}
\label{entC}
S_{\rm e} \approx -\frac{1}{\log_2 N} \int_{-1}^{1}dz\,  |\psi(z)|^2 \log_2\left( \frac{2}{N} |\psi(z)|^2 \right), 
\end{equation} 
it is easy to check
that in the Gaussian regimes
\begin{equation}
\label{entG}
S_{\rm e}(\gamma) = \frac{1}{2}+\frac{C(\gamma)}{2 \log_2 N} 
%\log_2\left(\frac{\pi N e}{2} \sigma_\gamma^2\right)
% \left\{
% \begin{array}{ll}
%   \log\left(\frac{\pi}{2}\right)& \gamma_1 \!\ll\! \gamma \!\ll\! \gamma_2 \\
% & \gamma_3 \!\ll\! \gamma \!\ll\! \gamma_4
% \end{array}
% \right.
\end{equation}
where 
%$\sigma_\gamma^2$ is given in Eq.~\eqref{sigma} and
\begin{equation}
\label{entG2}
C(\gamma) =
 \left\{
 \begin{array}{ll}
   \log_2\left(\frac{ \pi e}{2 |\gamma|\sqrt{\gamma^2-1}}\right) & \gamma_1 \!\ll\! \gamma \!\ll\! \gamma_2 \\
 \log_2\left(\frac{\pi e}{2\sqrt{\gamma+1}}\right) & \gamma_3 \!\ll\! \gamma \!\ll\! \gamma_4
 \end{array}
 \right.
\end{equation}

Note that, despite Eq.~\eqref{entent} forbids that $S_{\rm e}(\gamma)$ 
exceed unity, the quantity in Eq.~\eqref{entG}  diverges in the vicinity of the mean-field critical point. However, this is not alarming, since at any finite boson population there is a small interval around such point where the ground-state is not Gaussian and Eq.~\eqref{entG} does not apply.
It could be easily proven that the requirement that  Eq.~\eqref{entG} be less than unity provides an estimate for the boundaries of the non-Gaussian regime that is equivalent to the one derived in Section \ref{lov}.
An interesting feature revealed by Eqs.~\eqref{entG}-\eqref{entG2} is that  
 in the $N\to \infty $ limit the entanglement entropy tends to a constant, independent of the interaction to hopping ratio, $S_{\rm e}(\gamma) = \frac{1}{2}\;\forall\,\gamma \neq \gamma_{\rm c}$. Note indeed that quantity in Eq.~\eqref{entG2} does not depend on $N$, so that the second term in Eq.~\eqref{entG} vanishes with increasing population.
\begin{figure}
\includegraphics[width= \columnwidth]{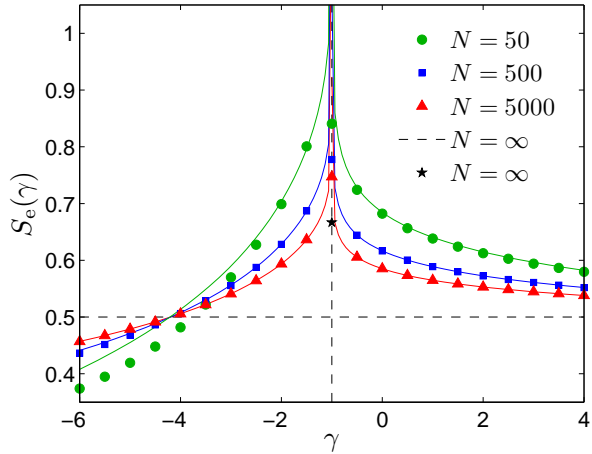}
\caption{\label{entF1} (color online) The analytic expression for the entanglement entropy, Eq.~\eqref{entG},  is compared with numerical results at different boson populations. Numerical data (symbols) and analytic results (solid lines) corresponding to the same boson population have the same color. The black dashed horizontal line is the limit of the Gaussian result,  Eq.~\eqref{entG}. The black star is the thermodynamic limit of the entanglement entropy at the  mean-field critical point, Eq.~\eqref{entCr}. The vertical dashed line signals the mean-field critical point.
}
\end{figure}

\begin{figure}
\includegraphics[width= \columnwidth]{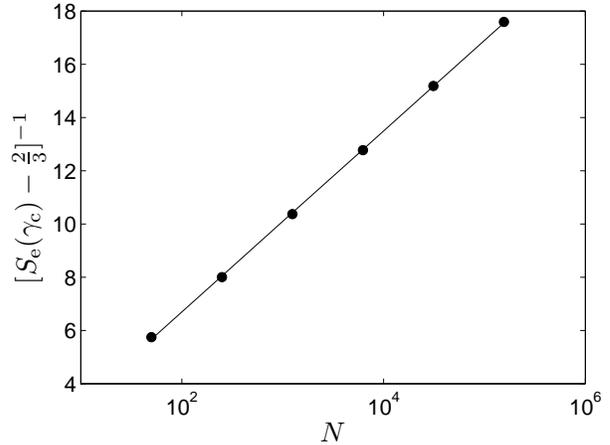}
\caption{\label{entF2} Scaling of the entanglement entropy at the
  mean-field critical value. The data points are obtained from exact
  diagonalization of system containing $N=50 \cdot 5^n$ bosons, with
  $n=0,\cdots,5$. The thin solid line corresponds to $3.4 \log_2 N -0.11$.
}
\end{figure}

The behavior of the entropy of entanglement in the critical regime can be obtained by plugging $\psi(z) = N^{1/6} \phi_0(N^{1/3} z)$ into Eq.~\eqref{entC}, where $\phi_0(\zeta)$ is once again the ground state of Eq.~\eqref{ceq4}.
 Straightforward calculations give
\begin{equation}
\label{entCr}
S_{\rm e}(\gamma) =\frac{2}{3} - \frac{1}{\log_2 N} \int d\zeta\, |\phi_0(\zeta)|^2 \log_2 \left[2 |\phi_0(\zeta)|^2\right]
\end{equation}
Since the integral in the second term of Eq.~\eqref{entCr} does not depend on $N$, we get that $\lim_{N\to \infty} S_{\rm e}(\gamma_{\rm c}) = \frac{2}{3}$.
Thus the mean-field critical point is signalled by the entropy of entanglement as a peak which, in the large population limit, turns into a  discontinuity.
Figure \ref{entF1} shows a comparison between  Eqs.~\eqref{entG}-\eqref{entG2} and numerical results obtained via exact diagonalization. The agreement is quite satisfactory, and improves with increasing boson population.
Note that even at relatively large populations the entropy of entanglement is still rather different from its limit. This is because the finite size corrections decrease logarithmically with $N$.
The different behavior at criticality is demonstrated in Fig~\ref{entF2}, which shows the correctness of Eq.~\eqref{entCr}.

\section{Thermal fluctuations}
\label{thermS}
We now turn to the thermal fluctuations of the order parameter $\hat z$
\begin{equation}
\label{therm1}
\langle \hat z^2 \rangle_T-\langle \hat z \rangle^2_T=\frac
{\sum_n e^{-\beta E_n} \left(\langle \hat z^2 \rangle_n-\langle \hat z \rangle^2_n\right)}
{\sum_n e^{-\beta E_n}}
\end{equation}
where $\langle \cdot \rangle_T$ and $\langle \cdot \rangle_n$ denotes the thermal average and the quantum expectation value on the state of energy $E_n$, respectivley. 

\begin{figure}
\includegraphics[width=8cm]{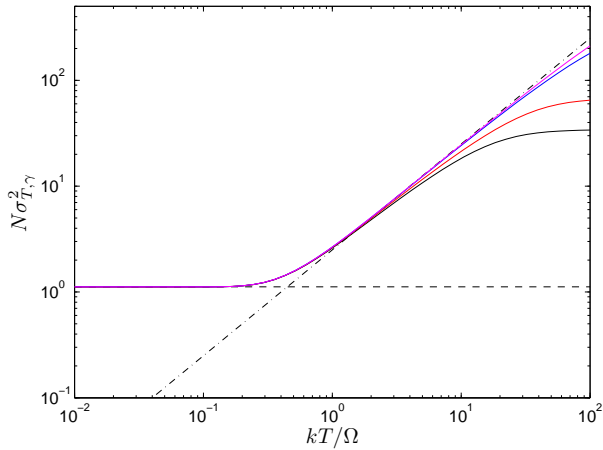}
\caption{\label{harmonic_kT} (color online) Scaling of imbalance fluctuations as a function of temperature for $\gamma=-0.2$. Black, red, blue and magenta solid line correspond to 100, 200, 1000 and 2000 bosons, respectively. Dashed  and dash-dotted line is the theoretical predictions \eqref{sigma} and \eqref{therm4}, respectively.}
\end{figure}

Evaluating $E_n$, $\langle \hat z^2 \rangle_n$ and $\langle \hat z \rangle_n$ using standard properties of quantum hamonic oscillators (see Section \ref{HarmAppr}) we get
\begin{equation}
\label{therm2}
\langle \hat z^2 \rangle_T-\langle \hat z \rangle^2_T= z_{\gamma}^2+\sigma_{T,\gamma}^2
\end{equation}
where the thermal fluctuations are 
\begin{equation}
\label{therm3}
\sigma_{T,\gamma}^2=\left\{
\begin{array}{ll}
\frac{\coth\left(\frac{\Omega\sqrt{\gamma^2-1}}{2k_{\rm B} T}\right)}
{N |\gamma| \sqrt{\gamma^2-1}}, & \gamma_1\ll \gamma \ll \gamma_2 \\[1em]
\frac{\coth\left(\frac{\Omega\sqrt{\gamma+1}}{2k_{\rm B} T}\right)}
{N \sqrt{\gamma+1}}, & \gamma_3\ll \gamma \ll \gamma_4
\end{array}
\right.
\end{equation}
and $z_\gamma$ is given in Table \eqref{coeff}.
Therefore for $k_{\rm B} T< \Delta E_2$ (i.e. the lowest energy gap of the system) $\sigma_{T,\gamma}^2$ is essentially given by the fluctations of the quantum ground states \eqref{ez2},  whereas for large temperatures the thermal fluctuation dominates and we have 
\begin{equation}
\label{therm4}
\sigma_{T,\gamma}^2\approx \left\{
\begin{array}{ll}
\frac{2k_{\rm B} T}
{\Omega N |\gamma| (\gamma^2-1)}, & \gamma_1\ll \gamma \ll \gamma_2 \\[1em]
\frac{2k_{\rm B} T}
{\Omega N (\gamma+1)}, & \gamma_3\ll \gamma \ll \gamma_4
\end{array}
\right.
\end{equation}
Eq. \eqref{therm4} highlights the typical linear dependence on temperature of the thermal fluctuations in harmonic systems. We remark that both thermal and quantum fluctuations decrease with the total population as $1/N$. This is demonstrated by the data collapse in Fig.~\ref{harmonic_kT}.

Also in the thermal regime we expect the harmonic approximation to apply only when the higher
order therms in Eq. \eqref{ceq2} are negligible. This means that for $\gamma_1<\gamma<\gamma_2$, $\sigma_{T,\gamma}^2$ should be smaller than $A/B$ while for $\gamma_3<\gamma<\gamma_4$, $\sigma_{T,\gamma}^2$ should be smaller than $A/C$. In both cases for $\gamma\approx -1$ we get $k T \lesssim N(\gamma+1)^2$. Note indeed that, as  evident from Fig.~\ref{harmonic_kT}, the higher the temperature, the larger must the population be for the   harmonic approximation (dash-dotted line) to apply.

\begin{figure}
\includegraphics[width=0.9 \columnwidth]{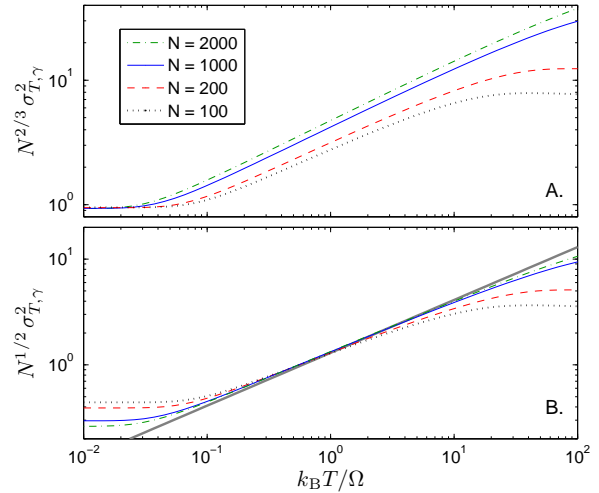}
\caption{\label{crit_kT} (color online) Scaling of imbalance fluctuations as a function of temperature for $\gamma=\gamma_{\rm c}=-1$ and different boson populations. The same data are multiplied by a different function of the boson population in the two panels, to highlight the different scaling laws  in the quantum and thermal regime. The thick gray line in panel B is the typical behaviour of the classical thermal fluctuations for a pure quartic potential, Eq.~\eqref{therm6}. }
\end{figure}

Let us now consider the regime $\gamma_2<\gamma<\gamma_3$ where the quartic term dominates. The values of $E_n$, $\langle \hat z \rangle_T$ and $\langle \hat z^2 \rangle_n$ can be estimated within the semiclassical Bohr-Sommerfeld approximation discussed in Sec. \ref{crS}. Plugging Eqs. \eqref{ceq6} and \eqref{ceq7} into Eq. \eqref{therm1} we get, for $\gamma=\gamma_{\rm c}=-1$,
\begin{equation}
\label{therm5}
\langle \hat z^2 \rangle_T-\langle \hat z \rangle^2_T\approx\sigma_{\gamma,T}\approx
\frac{\sum_n e^{-\frac {\beta C_1 \Omega}{N^{1 /3}} (n+\frac 1 2)^{\frac 4 3}} C_2 (n+ \frac 1 2)^{\frac 2 3}}
{N^{\frac 2 3}\sum_n e^{-\frac {\beta C_1 \Omega}{N^{1 /3}} (n+\frac 1 2)^{\frac 4 3}}}.
\end{equation}
It is easy to check that also in this critical regime $\sigma_{\gamma,T}$ reduces to the purely quantum contribution when $k_{\rm B} T$ is smaller than the energy gap between the lowest eigenpair, $C_1 \Omega /N^{1/3}$.
At higher temperatures we have
\begin{equation}
\label{therm6}
\sigma_{\gamma,T}\approx
\frac{\int e^{-\frac {\beta C_1 \Omega}{N^{1 /3}} n^{\frac 4 3}} C_2 n^{\frac 2 3} dn}
{N^{\frac 2 3}\int e^{-\frac {\beta C_1 \Omega}{N^{1 /3}} n^{\frac 4 3}} dn} \approx
\left ( \frac {k_{\rm B} T}{N\Omega} \right)^{\frac 1 2} ,
\end{equation}
which is the typical behaviour of classical thermal fluctuations when a pure quartic potential is present. We remark that within the critical region the fluctuations of the population imbalance due quantum effects scale as $N^{-2/3}$ with increasing population, while the thermal fluctuations scale as $N^{-1/2}$. This is shown in the upper and lower panel of  Fig.~\ref{crit_kT}, respectively.
\begin{figure}
\includegraphics[width= \columnwidth]{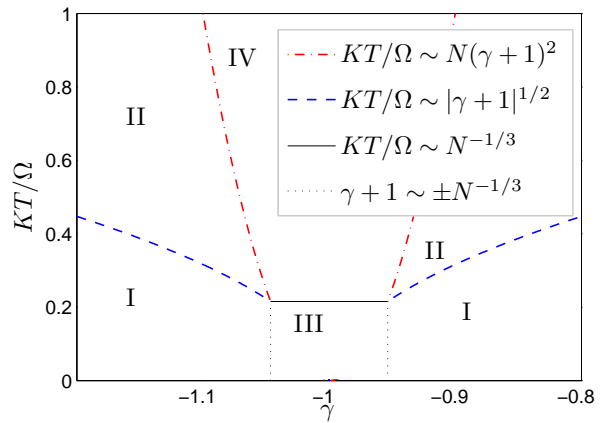}
\caption{\label{PhasDTemp} (color online) In this figure we represent the different thermal and quantum regimes in the region aroud the critical point $\gamma=-1$ the data are obtained for $N=100$. In regions I-IV  represent quantum harmonic, thermal harmonic, quantum quartic and thermal non-harmonic behaviours respectively.}
\end{figure}

The different quantum and classical regimes around $\gamma=-1$ can be summarized as in Figure \ref{PhasDTemp}. Note that the boundaries between different regions, signalled by the coloured lines, are actually  crossovers between different behviours. The only true critical point is $\gamma=-1$,  in the limit of infinite $N$.

 In regions I and III $\sigma^2_{\gamma,T}$ coincides with the result applying at zero-temperature for purely quartic and harmonic potentials, respectively. Region II is characterized by thermal fluctuations in a harmonic potential, while in region IV we have a classical behaviour in an anharmonic potential. In regions I and II $\sigma^2_{\gamma,T}$ decays as $N^{-1}$, while in region III quantum critical effects produce fluctuations decaying as $N^{-2/3}$. Finally, in region IV fluctuations should decay as $N^{-1/2}$ for not too large temperatures, and they should be independent of $N$  for very large $k_{\rm B} T/\Omega$.
\section{conclusions and perspectives}
\label{concl}
In this paper we addressed the quantum counterpart of the transition from a localized to a delocalized ground-state occurring in an interacting bosonic gas trapped in a double well potential, i.e. a bosonic Josephson junction. We identified and discussed five regimes depending on the value of the effective interaction among bosons. For sufficiently small effective interactions we analyzed the low-lying spectrum of the system Hamiltonian (a two-mode Bose-Hubbard model) by recasting the problem in terms of a Schr\"odinger-like equation for a single particle in a one-dimensional domain. In particular we studied how the results of the analysis on the quantum system reduce to those of the semiclassical treatment based on the Discrete Self-Trapping equations as the bosonic population is increased. We devoted a special attention to the critical regime which, in the theromodynamic limit of infinite bosonic population, collapses to a bifurcation point in the semiclassical picture. We extended our analysis to quantities borrowed from Quantum Information Theory, namely the entropy of entanglement and the ground-state fidelity, and to finite temperature effects.
In the latter respect, we evidenced that, in the critical regime, the quantum and thermal fluctuations of the population imbalance of the two wells exhibit different scaling behaviors, which may be amenable to quantitative measurement.
\acknowledgments
We acknowledge useful and stimulating discussions with Vittorio Penna.

\appendix

\section{Strong-coupling regimes}
\label{Apr}
As we mention in Section \ref{lov}, when interaction dominates over
kinetic energy the continuous approach of Section \ref{Sle} does not
apply. However, in this situation an approximation of the ground-state
of the system can be given by resorting to perturbative theory. 
For large repulsive interactions, first order perturbative theory
gives
\begin{equation}
\label{pertR}
%|\Psi\rangle =
%|\Psi_0\rangle+\frac{\Omega}{2U}\sqrt{\frac{N}{2}\left(\frac{N}{2}+1\right)}\left(|\Psi_+\rangle+|\Psi_-\rangle\right)
|\Psi\rangle =|\Psi_0\rangle+\lambda\left(|\Psi_+\rangle+|\Psi_-\rangle\right)
\end{equation}
where, assuming without loss of generality that $N$ is even,
\begin{equation}
\label{lamb}
\lambda = \frac{\Omega}{2U}\sqrt{\frac{N}{2}\left(\frac{N}{2}+1\right)}
\end{equation}
and
\begin{equation}
|\Psi_0\rangle=\left|\frac{N}{2},\frac{N}{2}\right\rangle,\quad
|\Psi_\pm\rangle=\left|\frac{N}{2}\pm 1,\frac{N}{2}\mp 1\right\rangle
\end{equation}
Equation \eqref{pertR} applies when the factor in front of the
first order perturbative term is small, $\lambda\ll 1$. It is easy to
check that, for this to be true, it should be $\gamma \gg \gamma_4$,
where $\gamma_4 \sim N^2$ is the upper bound of the ``single Gaussian''
regime discussed in Section \ref{lov}. Note that in passing from the
``single Gaussian'' regime, $\gamma_3 \ll \gamma \ll \gamma_4$, to the
present ``strong coupling'' regime, $\gamma \gg \gamma_4 $, there is a change in the behaviour of the fluctuations  
of the population imbalance operator, Eq.~\eqref{pio},
on the ground-state of the system, as  discussed in Ref. \cite{Javanainen_PRA_60_2351}. Indeed in the former regime this is
clearly $\langle\hat z^2\rangle-\langle\hat z\rangle^2 =
\sigma_\gamma^2$, where $\sigma_\gamma^2\sim N^{-1}$ is given in Eq. \eqref{sigma}.
% and we use the shorthand 
%\begin{equation}
%\langle\cdot\rangle = \frac{\langle\Psi|\cdot|\Psi\rangle}{\langle\Psi|\Psi\rangle}.
%\end{equation}
Conversely, in the strong coupling regime of perturbative state
\eqref{pertR}, we get
\begin{equation}
\label{fioR}
\langle\hat z^2\rangle-\langle\hat z\rangle^2 = \frac{8
  \lambda^2}{N^2(1+2\lambda^2)} \approx \frac{N^2}{2\,\gamma^2}
%\frac{\frac{\Omega^2}{U^2}\left(\frac{1}{2}+\frac{1}{N}\right)
%}{1+\frac{\Omega^2}{4U^2} N\left(\frac{N}{2}+1\right)}  \approx \frac{N^2}{2\,\gamma^2}
\end{equation}
where the last approximate equality applies in the limit of small
perturbations and large populations. 
Since for $\gamma \sim \gamma_4$ the fluctuations in Eqs. \eqref{sigma} and \eqref{fioR}
are of the same order of magnitude, there is no intermediate regime
between the ``single Gaussian'' and the repulsive ``strong-coupling''
regimes 

Similar arguments apply in the attractive ``strong-coupling'' regime. At first
order in perturbation theory the ground state is
\begin{equation}
\label{pertA}
|\Psi\rangle = |\Psi_0\rangle-\frac{\lambda}{\sqrt{2}}\left(|\Psi_+\rangle+|\Psi_-\rangle\right)
\end{equation}
where $|\Psi_0\rangle=\frac{1}{\sqrt{2}}\left(\left|N,0\right\rangle+\left|0,N\right\rangle\right)$,
%\begin{equation}
%|\Psi_0\rangle=\frac{\left|N,0\right\rangle+\left|0,N\right\rangle}{\sqrt{2}},
%\end{equation}
$|\Psi_+\rangle=\left|N-1,1\right\rangle$, $
|\Psi_-\rangle=\left|1,N-1\right\rangle$ and $\lambda = \frac{\Omega}{2U}\frac{\sqrt{N}}{N-1} $.
 The reqirement that  $\lambda \ll 1$ identifies the attractive
 ``strong-coupling'' regime $\gamma\ll \gamma_1 \sim
 -\sqrt{N}$. Straightforward calculations give
\begin{equation}
\label{fioS}
 \langle\hat z^2\rangle-\langle\hat z\rangle^2 
 =\frac{1+\lambda^2\frac{(N-2)^2}{N^2}}{1+\lambda^2} \approx  1-\frac{1}{\gamma^2}+ \frac{1}{\gamma^2 N}
\end{equation}
 We observe that, unlike the repulsive case, there is no
 cross-over of the fluctuations of operator \eqref{pio} in passing
 from the strong-coupling regime $\gamma\ll \gamma_1$ to the ``double
 Gaussian'' regime described in Section \ref{lov}. 
Note indeed that Eq. \eqref{fioS} coincides with Eq. \eqref{ez2} at large negative $\gamma$. The same results apply when one considers a single peak,
\begin{equation}
\label{pertAa}
|\Psi\rangle = |N,0\rangle+\lambda |N-1,1\rangle
\end{equation}
which gives
\begin{align}
\label{fio}
 \langle\hat z^2\rangle-\langle\hat z\rangle^2 &
 =\frac{1+\lambda^2\frac{(N-2)^2}{N^2}}{1+\lambda^2}- \left[\frac{1+\lambda^2\frac{N-2}{N}}{1+\lambda^2}\right]^2\nonumber\\
& \approx \frac{4 \lambda^2}{N^2} \approx \frac{1}{\gamma^2 N}
\end{align}
But this is exactly the same behavior as $\sigma_\gamma^2$ in Eq. \eqref{sigma} for large
populations and large (negative) $\gamma$.


\begin{thebibliography}{55}
\bibitem[{att()}]{attr_rep_map}
\bibinfo{note}{This occurs at sufficiently strong attractive interactions. As
  we discuss in the following, for a simple mapping shows that the same
  description applies to the highest-energy state of a repulsively interacting
  system case. Furthermore, the effects of an attractive interactions can be
  equivalently produced by a sufficiently strong repulsive interaction among
  bosons contained in different wells of the trapping potential.}
\bibitem[{int()}]{intN}
\bibinfo{note}{Rigorously, the integration domain is the interval $-1\leq z
  \leq 1$ but, since the Gaussian functions become strongly localized for large
  populations, it can be extended over the entire real line.}
\end{thebibliography}
\end{document}